\documentclass[12pt]{article}
\usepackage{amssymb,amsmath,amsthm,amsfonts,amscd}
\usepackage{graphicx}
\newcommand\be{\begin{equation}}
\newcommand\ee{\end{equation}}
\newcommand\bea{\begin{eqnarray}}
\newcommand\eea{\end{eqnarray}}

\newcommand{\fatalpha}{{\bf \alpha \kern -0.44em \alpha}}
\newcommand{\fatsigma}{{\bf \sigma \kern -0.54em \sigma}}
\newcommand{\tpchi}{{\bf \chi \kern -0.35em \chi}}
\newcommand{\llambda}{{\bf \lambda \kern -0.45em \lambda}}

\renewcommand{\theequation}{\arabic{equation}}
\renewcommand{\theequation}{\thesection.\arabic{equation}}
\bibliography{plain}
\title{\bf  Two and Three-Qubits Geometry, Quaternionic and Octonionic Conformal Maps, and Intertwining Stereographic Projection}

\author{ G. Najarbashi $^{a}$
 \thanks{Najarbashi@uma.ac.ir} ,B. Seifi $^{a}$ \thanks{B.seifi@uma.ac.ir } , S. Mirzaei
 $^{a}$ \thanks{Smirzaei@uma.ac.ir} \\
\\ \\
$^a${\small Department of Physics, University of Mohaghegh Ardabili, P.O. Box 179, Ardabil, Iran.}\\
{\small}} \pagebreak

\begin{document}
\maketitle
\newpage 

\begin{abstract}
In this paper the geometry of  two and three-qubit states  under local unitary groups is discussed. We first review the one qubit geometry and its relation with Riemannian sphere under the action of group $SU(2)$. We show that the  quaternionic stereographic projection intertwines between  local unitary group  $SU(2)\otimes SU(2)$ and  quaternionic
M\"{o}bius transformation. The invariant term appearing in this operation is related to concurrence  measure. Yet, there exists  the same intertwining stereographic projection  for much more global group $Sp(2)$, generalizing the familiar Bloch sphere in 2-level systems. Subsequently, we introduce  octonionic stereographic projection and octonionic conformal map (or octonionic M\"{o}bius maps) for three-qubit states and find evidence that they may have invariant terms under local unitary operations which shows that both maps are entanglement sensitive.

 {\bf PACs Index: 03.67.a 03.65.Ud}
\end{abstract}
\newpage
\section{Introduction}
There are  relations between geometry and structure of spinors in various area of quantum information theory \cite{Varadarajan,carol,da, stv, pla, plb, plc, kpam, ha,  kz}. The simplest quantum states involve qubits (two-level systems), including spin states of a spin $\frac{1}{2}$ particle, the polarization states of a photon, or the ground and excited state of an atom or ion.  Single qubit  have simple geometric picture, i.e. pure states  can be identified with points on the surface of the  Bloch sphere $S^{2}$ and mixed states identified with points inside the Bloch sphere. Therefore it is tempting to find geometric pictures for the
higher dimension quantum states which   resemble one qubit representation. Finding the set of the Bloch vector representation for N-level systems, generalizing the familiar Bloch vector in 2-level systems, seems to be nontrivial task \cite{gka, gki}.
\par
The relation between the  Hopf fibration, single qubit and two-qubit states has  been studied by Mosseri and Dandoloff \cite{rrd}  in quaternionic skew-field and subsequently have been generalized to three-qubit state based on octonions by Bernevig and Chen \cite{phdm} . However, there is also one more reason to look for  conformal maps. For two qubit states the concurrence measure appears explicitly in quaternionic stereographic projection  which geometrically means that non-entangled states are mapped from $S^7$ onto a 2-dimensional planar subspace of the target space ${\mathbb{R}}^{4}$. On the other hand it was shown that   third Hopf fibration is also entanglement sensitive for three qubit states \cite{phdm}.
\par
However, it seems that there is also another geometrical approach to describe  one two and three qubit
states called M\"{o}bius transformation \cite{g.n, Lee}. As is typical in physics, the local properties are
more immediately useful than the global properties, and the local unitary transformation is
of great importance. Therefore in this paper, we pursue a different approach to study the
geometrical structure of two and three-qubit states under a local unitary transformation \cite{sdgc}. We
show that the quaternionic stereographic projection  of two-qubit states intertwines
between the local unitary $SU(2)\otimes SU(2)$ and corresponding quaternionic M\"{o}bius transformations  \cite{aslaken, harvey}, which can be useful in theoretical physics such as quaternionic quantum
mechanics \cite{adler}, quantum conformal field theory \cite{sdgc,aslaken,harvey, adler, dpmp}  and quaternionic computations
\cite{fjm}. This generalizes our early work  restricted to group $SO(2)\otimes SU(2)$ \cite{g.n}.
However the action of transformations that involve with non-commutative quaternionic
skew-field on a spinor (living in quaternionic Hilbert spaces) is more complicated than the
complex one. Roughly speaking one must distinguish between the left and right actions of
a quaternionic transformations on a given state  \cite{lsdc}. This anomalous property of
quaternionic transformation lead us of defining the special quaternionic M\"{o}bius transformations.
An additional goal of this work is to generalize all feature to three-qubit states.
While the formulation is  almost trivial task for two-qubit case, the problem is more involved in all features
 as we must cast the three qubit states in noncommutative and non-associative octonion skew-field.
we will show that the construction of octonionic  stereographic projection  and M\"{o}bius transformation under local unitary group
$SU(2)\otimes SU(2)\otimes SU(2)$ turns out to be possible and as
in  two-qubit case, both  are  entanglement  sensitive \cite{epr, peres, horo1,  horo2,  horo3,  horo4}, i.e. there are terms that is invariant under local unitary transformation which is related to concurrence measure. In all of these construction, we will insist on commutativity of diagrams which  any two compositions of maps starting at one point in
the diagram and ending at another are equal. This will be entirely in the language of  algebraic topology \cite{heydari1, heydari2,hacher}.
\par
The paper is organized as follows. In section 2, we briefly summarize one-qubit geometry
and conformal map in a commutative diagram and give an example for  evolution of quantum system in the complex plane by Hadamard like transformation. In section 3, we extend  the results of Ref \cite{g.n} to general local and global  transformation. In section 4, we introduce the basic geometrical structure together with basic background material, incorporating all information we need for characterization of three-qubit geometry. Formulation the octonionic  stereographic projection and M\"{o}bius transformation under local unitary group $SU(2)\otimes SU(2)\otimes SU(2)$ is established in this section.  The paper is ended with a brief conclusion and two appendices.
\section{One-qubit geometry}
Let ${\mathcal{H}}_{d}^{\mathbb{F}}$ be a  $d$ dimensional Hilbert space  in the field or skew-field $\mathbb{F}=\mathbb{C}$, $\mathbb{Q}$ or $\mathbb{O}$ where $\mathbb{C}, \mathbb{Q}$ and  $\mathbb{O}$ are complex, quaternion and octonion field (or skew-field) respectively. An arbitrary one-qubit pure state in the complex two dimensional Hilbert space ${\mathcal{H}}_{2}^{\mathbb{C}}$, is given by
\begin{equation}\label{pure1}
|\psi\rangle=\alpha_{0}|0\rangle+\alpha_{1}|1\rangle\quad\quad,\quad
|\alpha_{0}|^2+|\alpha_{1}|^2=1\quad\quad,\quad
\alpha_{0},\alpha_{1}\in \mathbb{C}.
\end{equation}
We summarize the results of Ref. \cite{Lee} for one qubit pure state in a compact form as a  commutative diagram
\[\begin{CD}
{\mathcal{H}}_{2}^{\mathbb{C}}         @>{\mathcal{P}}>>      \mathbb{\widetilde{C}} \\
@V{A}VV               @VV{\mathcal{F}_{_{A}}}V \\
{\mathcal{H}}_{2}^{\mathbb{C}}         @>>{\mathcal{P}}>
\mathbb{\widetilde{C}}
\end{CD}\]
where $\mathcal{P}$ denotes the  stereographic projection for one-qubit state (\ref{pure1}),
i.e.
\begin{equation}
\mathcal{P}(|\psi\rangle):=\alpha_{0}\alpha_{1}^{-1}\in
\mathbb{\widetilde{C}}= \mathbb{C} \cup \{\infty\},
\end{equation}
and $\mathcal{F}_{_{A}}\in PSU(2)=SU(2)/\{\pm I\}$ is M\"{o}bius transformation corresponding to $2\times2$ matrix $A\in SU(2)$   define as
\begin{equation}
A=\left(\begin{array}{cc} a& b
\\-\bar{b}&\bar{a}
\end{array}\right)\ \longleftrightarrow\ \mathcal{F}_{_{A}}(z):=\frac{az+b}{-\bar{b}z+\bar{a}},\quad\quad |a|^{2}+ |b|^{2}=1,\quad\quad a,b,z \in\mathbb{C}.
\end{equation}
The M\"{o}bius transformations, generating the conformal group in the complex
plane,   can be identified  with
conformal transformations on the sphere using stereographic projection. Commutativity of the above diagram means
that for any one-qubit state $|\psi\rangle$ and any $A\in SU(2)$
we have
\begin{equation}
\mathcal{F}_{_{A}}\mathcal{P}(|\psi\rangle)=\mathcal{P}(A|\psi\rangle).
\end{equation}
\begin{figure}
\centerline{\includegraphics[width=10cm]{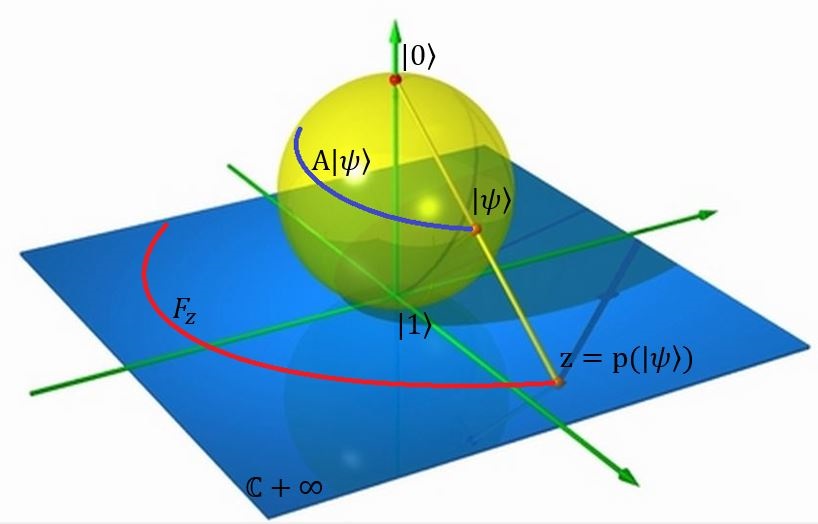}}
\caption{\label{appendix Fig.1}\small {Stereographic map of Bloch sphere and   time evolution of quantum states} \label{appendix Fig.1} }
\end{figure}
This shows that the  stereographic projection $\mathcal{P}$ intertwines
between any single qubit unitary operation $A$ and it's
corresponding M\"{o}bius transformation $\mathcal{F}_{_{A}}$. The basic idea behind the stereographic projection $\mathcal{P}$  is easily to project the point on Bloch sphere (a pure quantum state) onto the point on the complex plain (see Fig. \ref{appendix Fig.1}).
The pure state (\ref{pure1}) corresponds to  the spin $\frac{1}{2}$ quantum states and it is well-known that it's time
evolution is governed by  $2\times2$  unitary matrices.
For example, when we compare
the states that differ by an exchange of coordinates on Riemannian sphere, we should, at least in principle, be able to tell by what unitary operation we effect this exchange for otherwise, we cannot really compare them
other than in a formal and, in fact, in an ambiguous sense. According  to Schr\"{o}dinger equation
$
i\hbar \frac{\partial }{{\partial t}}\psi(x,t)  = \mathcal{\hat H}\psi(x,t),
$
 all information about any quantum mechanical system is contained in the matrix
elements of its time evolution operator
\begin{equation}
\psi(x,t)  =e^{- i\mathcal{\hat H }t}\psi(x,0),
\end{equation}
where $e^{- i\mathcal{\hat H} t}\equiv A\in SU(2)$, up to a global phase, describes dynamical evolution under the influence of the Hamiltonian from a time zero to time t.
For example if we take the  Hamiltonian as Hadamard  gate
\begin{equation}
H = \frac{1}{{\sqrt 2 }}\left( {\begin{array}{*{20}{c}}
  1&1 \\
  1&{ - 1}
\end{array}} \right),
\end{equation}
 and  initial state as  $|1\rangle$, then its time evolution on Bloch sphere reads
\begin{figure}
\centerline{\includegraphics[width=10cm]{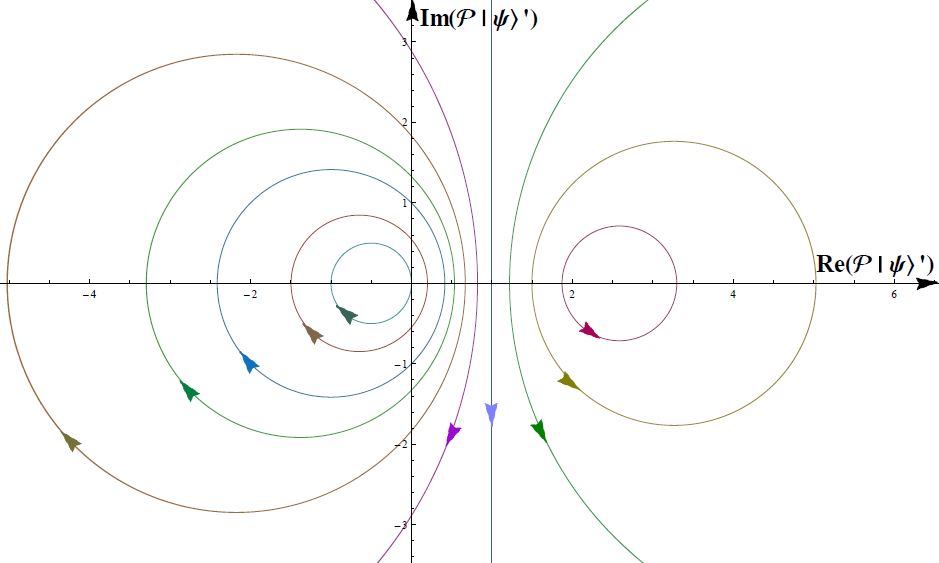}}
\caption{\label{appendix Fig.2}\small { Time evolution of the state  $(\cos\frac{\theta}{2}|0\rangle+e^{i\phi}\sin\frac{\theta}{2}|1\rangle )$ for various   initial states with $\phi=0$ and $\theta=\frac{3}{15}, \frac{4}{15}, ..., \frac{15}{15}$ by Hadamard like Hamiltonian.} \label{appendix Fig.2} }
\end{figure}
\begin{equation}\label{exam1}
|\psi(t)\rangle=e^{-iHt}|1\rangle=-\frac{i}{\sqrt{2}}\sin(t)|0\rangle+(\cos(t)+\frac{i}{\sqrt{2}}\sin(t))|1\rangle,
\end{equation}
which in turn implies that its corresponding time dependent stereographic projection is
\begin{equation}\label{eexam1}
\mathcal{P}(e^{-iHt}|1\rangle)=\frac{-1}{2 \cos^{2}(t)+\sin^{2}(t)}( \sin^{2}(t)+\sqrt{2}i\ \sin(t) \cos(t)),
\end{equation}
(see Fig \ref{appendix Fig.3}).
On the other hand if  we first  map the initial state to complex plane $\mathcal{P}|1\rangle=0$ and then take its
M\"{o}bius transformation
\begin{equation}
\mathcal{F}_{_{H}}(e^{-iHt}|1\rangle)=\frac{ sin(t)}{\sqrt{2}  i\ cos(t) - sin(t)},
\end{equation}
which one can easily verified that it is equal to the right hand side of Eq. (\ref{eexam1}). An important geometrical property of M\"{o}bius transformation is that, they map circles onto circles as depicted in Fig (\ref{appendix Fig.2}). This also includes straight lines as circles of infinite radius.  This result should not be too surprising:  There is a group theoretical identity behind this observation, namely the isomorphism $SU(2)\simeq S^2 \times U(1)$,
where $U(1)$ is the gauge subgroup contained in $SU(2)$. Indeed, starting with any initial state and  Hamiltonian yields the same result as  above which come from the fact that the  M\"{o}bius transformations preserve the angles (hence the
name conformal).
\section{Two-qubit geometry}
It is tempting to try to extend the results of the last section  to the system of bipartite two-qubit systems. However due to the difference between the dimensions of single qubit and two-qubit systems, there are a number mathematical differences which make the structure of the problem   a much harder (but also more interesting!). Firstly, in one qubit quantum systems
the state $\rho$ can be represented by its Bloch sphere representation, a vector $\vec r$ inside the unit ball in $\mathbb{R}^{3}$. The correspondance is given by
\begin{equation}\label{bloch}
\rho=\frac{1}{2}(\mathbb{I}+\vec r.\vec\sigma) ,\quad\quad r_{i}=tr(\rho\sigma_{i}),
\end{equation}
where $\mathbb{I}$ is identity matrix, $\{\sigma_{i}\}$ are the Pauli matrices and the geometry of state $\rho$ is completely described by the vector  $\vec r$ on Bloch sphere such that, the points on the surface of the Bloch sphere correspond to the pure states of the system, whereas the interior points correspond to the mixed states. In general there is nothing like  Bloch sphere representation for general two-qubit mixed states \cite{gka,gki}  but rather one is dealing with a particular quaternionic representation for  two-qubit pure states.
Second, for two-qubit system there is a quantum  correlation, i.e. entanglement which depends on locally or globally unitary transformation, therefore we must distinguish between local and global transformation.
\subsection{Local unitary transformations}
The Hilbert space ${\mathcal{H}}_{4}^{\mathbb{C}}$ of the
compound system is the tensor product of the individual Hilbert
spaces ${\mathcal{H}}_{2}^{\mathbb{C}}\otimes{\mathcal{H}}_{2}^{\mathbb{C}}$ with a direct product basis $\{|00\rangle,|01\rangle,|10\rangle,|11\rangle\}$. Any two-qubit pure state in this basis reads
\begin{equation}\label{twoqubit}
|\psi\rangle=\alpha|00\rangle+\beta|01\rangle+
\gamma|10\rangle+\delta|11\rangle\quad\quad\quad
\alpha,\beta,\gamma,\delta,\in \mathbb{C},
\end{equation}
with normalization condition $|\alpha|^2+|\beta|^{2}+|\gamma|^{2}+|\delta|^{2}=1$.
We summarize the result of \cite{g.n} in a commutative diagram fashion covenant for our purposes as
\[\begin{CD}
{\mathcal{H}}_{4}^{\mathbb{C}}         @>{\mathcal{Q}}>>      {\mathcal{H}}_{2}^{\mathbb{Q}}    @>{\mathcal{P}}>> \mathbb{\widetilde{Q}} \\
@V{B}VV               @V{\mathbb{Q}B}VV           @VV{\mathcal{F}_{_{B}}}V\\
{\mathcal{H}}_{4}^{\mathbb{C}}       @>{\mathcal{Q}}>>
{\mathcal{H}}_{2}^{\mathbb{Q}}    @>{\mathcal{P}}>>
\mathbb{\widetilde{Q}}
\end{CD}\]
where $\mathcal{Q}$ denotes the quaternification operation which maps every $|\psi\rangle\in {\mathcal{H}}_{4}^{\mathbb{C}}$  to quaterbit  $|{\psi}\rangle_{q}\in
{\mathcal{H}}_{2}^{\mathbb{Q}}$ as follows
\begin{equation}
\mathcal{Q}(|\psi\rangle):=|{\psi}\rangle_{q}=q_{1}|{0}\rangle_{q}+q_{2}|{1}\rangle_{q}=\left(
                                                                                          \begin{array}{c}
                                                                                            q_{1} \\
                                                                                            q_{2} \\
                                                                                          \end{array}
                                                                                        \right)
,\quad \quad |q_{1}|^2+|q_{2}|^2=1,
\end{equation}
where $q_{1}=\alpha+\beta e_{2}$ and $q_{2}=\gamma+\delta
 e_{2}$ are quaternion numbers (see appendix A). Note that this map is bijective (one to one and onto).  $\mathcal{P}$ is the quaternionic stereographic projection for quaterbit  defined as
\begin{equation}\label{confq}
\mathcal{P}(|\psi\rangle_{q}):=q_{1}q^{-1}_{2}=\frac{1}{|q_{2}|^2 }(S+Ce_{2}) \in
\mathbb{\widetilde{Q}}= \mathbb{Q} \cup \{\infty\},
\end{equation}
where $S=\alpha\overline{\gamma}+\beta\overline{\delta}$  and $C=\beta\gamma-\alpha\delta$ are Schmidt and concurrence terms respectively. If $S=0$, then the two-qubit pure state have Schmidt form $(\sqrt{\lambda_{1}}|00\rangle+\sqrt{\lambda_{2}}|11\rangle)$ with positive numbers $\lambda_{1}$ and $\lambda_{2}$. On the other hand  $2|C|$ is concurrence measure for two-qubit pure state meaning that if $C=0$ then the two qubit pure state (\ref{twoqubit}) is reduced to a separable state, i.e. factorized as $|\psi\rangle_{AB}=|\psi\rangle_{A}|\psi\rangle_{B}$ \cite{wootters}. This means that there is an one to one correspondence between the points on complex plane $\mathbb{\widetilde{C}}$ and two-qubit separable states.
 Let us mention that that the map $\mathcal{P}$ is related to the stereographic projection of the second Hopf fibration of the form
 \begin{equation}
\mathcal{P}:  \mathbb{Q}^{2} \xrightarrow{}   \mathbb{Q}\mathbb{P}^{1},
\end{equation}
where $\mathbb{Q}\mathbb{P}^{1}$ is the one dimensional quaternionic projective space. In this diagram $B\in SU(2)\otimes SU(2)$ is any local unitary transformation acting on two-qubit pure state (\ref{twoqubit})
\begin{equation}\label{trans1}
 |\psi\rangle'=A \otimes A'|\psi\rangle, \ \ \  \ A,A'\in SU(2).
\end{equation}
Remembering that, $SU(2)$ transformation can be parameterized in the complex field as
\begin{equation}
A=\left( {\begin{array}{*{20}{c}} a& b \\  { - \bar b}& \bar{a}
\end{array}} \right),\quad\quad |a^2|+|b^2|=1,
\end{equation}
we may alternatively write the transformation (\ref{trans1}) in quaternionic   right module as follows
\begin{equation}\label{module}
|{\psi}\rangle' _{q}=\mathbb{Q}B(|{\psi}\rangle_{q}):= A | {\psi}\rangle_{q}{\mathcal{A}'}^{(q)}= \left( {\begin{array}{*{20}{c}} a& b \\  { -\bar b}&\bar{a}
\end{array}} \right)\left(\begin{array}{c}
                      q_{1} \\
                      q_{2}
                    \end{array}\right)(a'-\bar b' e_{2})
,
\end{equation}
where
\begin{equation}\label{iso1}
{\mathcal{A}'}^{(q)}=a'-\bar b' e_{2}
\end{equation}
and the second equality by itself defines the $\mathbb{Q}B$ operation in the diagram. The definition  ${\mathcal{A}'}^{(q)}=a'-\bar b' e_{2}$ as quaternionic representation for unitary matrix $A'$ come from the isomorphism $Sp(1)\simeq SU(2)$.
Because of the associativity of quaternion numbers, there is no ambiguity when it comes to forming products of higher order, i.e. $(q_{1}q_{2})q_{3}=q_{1}(q_{2}q_{3})=q_{1}q_{2}q_{3}$.
 Finally, the quaternionic M\"{o}bius transformation $\mathcal{F}_{_{B}}$ associated to the local unitary transformation
 $B=A\otimes A' \in SU(2)\otimes SU(2)$ is defined as
 \begin{align}\label{cmt}
&\mathcal{F}_{_{B}}(q):= [{{aq{\mathcal{A}'}^{(q)} + b{\mathcal{A}'}^{(q)}}}][{{ - \bar bq{\mathcal{A}'}^{(q)} + \bar a{\mathcal{A}'}^{(q)}}}]^{-1}\nonumber\\
&\ \ \ \ \ \ \ \ \   =\frac{S'+C'e_{2}}{|q_{2}|^{2}|b|^{2}+|a|^{2}-2Re(\bar b q_{2}a)}
\end{align}
where $q=\mathcal{P}(|{\psi}\rangle_{q})$ is stereographic projection of the initial state (\ref{twoqubit}) and $Re(q)=\frac{1}{2}(q+\bar{q})$ is a real part of quaternion number $q$ and  $C'=C=(\beta\gamma-\alpha\delta)$ is  concurrence term which  is invariant under local unitary operations, as one would expect, while
\begin{equation}
S'=(|q_{2}|^{2}-|q_{1}|^{2})ab+S|a|^{2}-\bar S |b|^{2}
\end{equation}
is Schmidt term which unlike the concurrence term is not invariant under local unitary groups, even if the initial two-qubit state (\ref{twoqubit}) being  in Schmidt form in the old basis, i.e. $S=0$. Summarizing, we have found that the quaternionic stereographic projection $\mathcal{P}$ intertwines between the unitary local operation $B$
 and the corresponding quaternionic M\"{o}bius transformation $\mathcal{F}_{_{B}}$ viz.
\begin{equation}\label{ww}
\mathcal{P}\mathcal{Q}B|\psi\rangle
=\mathcal{P}(\mathbb{Q}B)\mathcal{Q}|\psi\rangle
=\mathcal{F}_{_{B}}\mathcal{P}\mathcal{Q}|\psi\rangle.
\end{equation}
It should be mentioned that  $\mathcal{Q}B$ acting on quaterbit $|\psi\rangle_{q}$, endowed with right module as Eq. (\ref{module}), belongs to the general form of local unitary transformation and there was no need to do  with  restriction $SO(2)\otimes SU(2)$ or
 $SU(2)\otimes SO(2)$ as established in Ref. \cite{g.n} which, in turn, implies that the imaginary part of stereographic projection (\ref{confq}) is also an entanglement measure for two-qubit pure states.
\subsection{Global unitary transformations Sp(2)}
In deriving the result (\ref{ww}) we have restricted ourselves to  the discussion of local unitary transformations. For less restrictive geometry we consider quaternionic spinor $|\psi\rangle_{q}=q_{1}|0\rangle_{q}+q_{2}|1\rangle_{q}$
and demand its transformation under global  unitary transformations $Sp(2)$,
the quaternionic counterpart of group $SU(2)$, which is defined as
\begin{equation}
Sp(2):= \{A\in GL(2,\mathbb{Q}) : A^{\dagger}A=I\},
\end{equation}
or equivalently  it can be expressed as
\begin{equation}
Sp(2):= \{U\in U(4) : UJU^T=J\},
\end{equation}
where $J= I\otimes (-i\sigma_{2})$ and $\sigma_{2}$ is usual second Pauli  matrix.
If we define the action of every element of  $Sp(2)$ on quaternionic spinor as
\begin{equation}
|\psi\rangle'_{q}=A|\psi\rangle_{q}=\left(
\begin{array}{cc}
 p_1 & p_2 \\
 p_3 & p_4 \\
\end{array}
\right) \left(
\begin{array}{c}
 q_1 \\
 q_2 \\
\end{array}
\right)=\left(
\begin{array}{c}
 p_1 q_1+p_2 q_2 \\
 p_3 q_1+p_4 q_2 \\
\end{array}
\right),
\end{equation}
 and its corresponding  M\"{o}bius transformation as follows
\begin{equation}\label{mob1}
\mathcal{F}_{{_A}}(q)=[p_{1}q+p_{2}][p_{3}q+p_{4}]^{-1},
\end{equation}
with all $p_{i}, q_{i}\in \mathbb{Q}$ being quaternionic number, then the stereographic projection (\ref{confq}) intertwines between   $Sp(2)$
  and M\"{o}bius transformation, i.e. $\mathcal{P}(A|\psi\rangle)=\mathcal{F}_{_{A}}\mathcal{P}|\psi\rangle$.
Written more explicitly,
\begin{equation}\label{form1}
\mathcal{P}(Sp(2)|\psi\rangle_{q})=\mathcal{F}_{_{Sp(2)}}(\mathcal{P}|\psi\rangle_{q})=\frac{(|q_{1}|^{2}p_{1}\bar p_{3}+p_{1}q_{1}\bar q_{2}\bar q_{4}+p_{2}q_{2}\bar q_{1}\bar q_{3}+|q_{2}|^{2}p_{2}\bar p_{4})}{| p_3 q_1+p_4 q_2|^{2}},
\end{equation}
or in the  diagrammatic form actually reads
\[\begin{CD}
{\mathcal{H}}_{2}^{\mathbb{Q}}         @>{\mathcal{P}}>>      \mathbb{\widetilde{Q}} \\
@V{Sp(2)}VV               @VV{\mathcal{F}_{_{Sp(2)}}}V \\
{\mathcal{H}}_{2}^{\mathbb{Q}}         @>>{\mathcal{P}}>
\mathbb{\widetilde{Q}}
\end{CD}\]
The derivation made here is based on  explicit use of the particular form
of  M\"{o}bius transformation (\ref{mob1}) and as in the case of local unitary transformation it is unique.
The description above, albeit perfectly valid, still suffers from a deficiency: There is   two-qubit state representation, like as  Eq. (\ref{twoqubit}), for quaternionic spinor but the first equality in Eq. (\ref{ww}) diagram does not hold for quaternionic spinor and the discussion may have seemed somewhat abstract. Nonetheless, the derivation is instructive and helps to understand the algebraic topology of quaternionic spinors (see, e.g. \cite{hacher}).
\section{Geometry of three-qubit states under local unitary groups}
This section devoted to provide some basic tools and background to describe the geometry of three-qubit pure states under local unitary transformations.
\subsection{Octonionic conformal map}
The Hilbert space ${\mathcal{H}}_{8}^{\mathbb{C}}$ for  three-qubit system is the tensor product of the individual Hilbert spaces ${\mathcal{H}}_{2}^{\mathbb{C}}\otimes{\mathcal{H}}_{2}^{\mathbb{C}}\otimes{\mathcal{H}}_{2}^{\mathbb{C}}$ with a direct product basis $\{|000\rangle,|001\rangle,|010\rangle, 011\rangle ,|100\rangle,|101\rangle,|110\rangle,|111\rangle\}$. An arbitrary three-qubit pure state in this basis read as
\begin{equation}\label{threequbit}
|\psi\rangle=t_{0}|000\rangle+t_{1}|001\rangle+
t_{2}|010\rangle+t_{3}|011\rangle+t_{4}|100\rangle+t_{5}|101\rangle+
t_{6}|110\rangle+t_{7}|111\rangle
,
\end{equation}
where $t_{i}\in \mathbb{C}$ are complex numbers and satisfy the normalization condition
$\sum_{i=0}^7|t_{i}|^2=1$.
Once again, projecting onto a basis in which all coefficients are singly quaternionic number, we can equivalently rewrite every $|\psi\rangle\in {\mathcal{H}}_{8}^{\mathbb{C}}$ by two-quaterbit $|{\psi}\rangle_{q}\in {\mathcal{H}}_{4} ^{\mathbb{Q}}$ as
\begin{equation}\label{quater1}
\mathcal{Q}(|\psi\rangle):=|{\psi}\rangle_{q}=q_{1}|{00}\rangle_{q}+q_{2}|{01}\rangle_{q} +q_{3}|{10}\rangle_{q} +q_{4}|{11}\rangle_{q}
\end{equation}
where
\begin{equation}
q_{1}=t_{0}+t_{1}e_{2}\quad,q_{2}=t_{2}+t_{3}e_{2}\quad,q_{3}=t_{4}+t_{5} e_{2}\quad,q_{4}=t_{6}+t_{7}e_{2},\quad
\end{equation}
and $\sum_{i=1}^4|q_{i}|^2=1$, is normalization condition in the language of quaternion numbers.
By a straightforward rearrangement of the quaternion numbers, this can be rewritten as
\begin{equation}\label{octabit}
|{ \psi }\rangle_{o}=\mathcal{O}(\mathcal{Q}(|\psi\rangle))=\mathcal{O}|{\psi}\rangle_{q}=o_{1}|{ 0 }\rangle_{o}+o_{2}|{ 1 }\rangle_{o},
\end{equation}
where $o_{1}=q_{1}+q_{2}e_{4}$ and $o_{2}=q_{3}+q_{4}e_{4}$ are octonion numbers (see Appendix A).
Clearly, $|{ \psi }\rangle_{o}$ is an  octonion bit (octobit) which  belongs to the Hilbert space ${\mathcal{H}}_{2}^{\mathbb{O}}$.
Consider now a general octonion number
\begin{equation}
 o=x_{0}e_{0}+x_{1}e_{1}+x_{2}e_{2}+x_{3}e_{3}+x_{4}e_{4} +x_{5}e_{5}+ x_{6}e_{6}+x_{7}e_{7}, \quad x_{i}\in \mathbb{R},
\end{equation}
with  inverse $o^{-1}=\frac{1}{|o|^2}\bar{o}$ where $\bar{o}$ is the complex conjugate of octonion $o$
 defined as
\begin{equation}\label{bar}
 \bar{o}=x_{0}e_{0}-x_{1}e_{1}-x_{2}e_{2}-x_{3}e_{3}-x_{4}e_{4} -x_{5}e_{5}- x_{6}e_{6}-x_{7}e_{7}.
\end{equation}
  Another definition which is necessary to formulate the octonionic stereographic projection is to define  $\tilde o$ as
\begin{equation}\label{tild}
\tilde o=x_{0}e_{0}+ x_{1}e_{1} +x_{2}e_{2}+x_{3}e_{3}+x_{4}e_{4}+x_{5}e_{5}+x_{6}e_{6} -x_{7}e_{7}.
\end{equation}
Using the above definition and Eq. (\ref{octabit}) and  (\ref{bar}), the octonionic stereographic projection $\mathcal{P}$ is defined  \cite{rrd,phdm}
\begin{equation}\label{confo2}
\mathcal{P}(|{\psi}\rangle_{o}):=\tilde o_{1}\tilde o^{-1}_{2}=\frac{1}{|o_{2}|^2 }(S_{0}+S_{1}e_{2}+(S_{2}+S_{3}e_{2})e_{4}) \in
\mathbb{\widetilde{O}}= \mathbb{O} \cup \{\infty\},
\end{equation}
where
\begin{align}\label{cmot}
\begin{array}{c}
  {S_0} = t_{0}\bar{t_{4}}+ t_{1}\bar{t_{5}}+ t_{2}\bar{t_{6}}+ t_{3}\bar{t_{7}},\\
  {S_1} = t_{1}t_{4} - t_{0}t_{5}+\bar{t_{3}}\bar{t_{6}}-\bar{t_{2}}\bar{t_{7}},\\
  {S_2} =  t_{2}t_{4} - t_{0}t_{6}+\bar{t_{1}}\bar{t_{7}}-\bar{t_{3}}\bar{t_{5}},\\
  {S_3} = t_{2}t_{5} - t_{1}t_{6}+\bar{t_{3}}\bar{t_{4}}-\bar{t_{0}}\bar{t_{7}},
\end{array}
\end{align}
and  $|\tilde o_{2}|^{2}=|o_{2}|^{2}=|t_{4}|^{2}+|t_{5}|^{2}+|t_{6}|^{2}+|t_{7}|^{2}$. An alternative, and often more useful  representation of $\tilde o_{1}$ and $\tilde o_{2}$  is given by
\begin{align}
\begin{array}{c}
  \tilde o_{1}=t_{0}+t_{1}e_{2}+(t_{2}+{\bar t}_{3}e_{2})e_{4}, \\
  \tilde o_{2}=t_{4}+t_{5}e_{2}+(t_{6}+{\bar t}_{7}e_{2})e_{4}.
\end{array}
\end{align}
 Note that, due to the non-associativity of octonions, i.e. $o_{1}(o_{2}o_{3})\neq(o_{1}o_{2})o_{3}$ one must keep the parenthesis in all equations involving the product of three or more octonions.
The octonionic stereographic projection $\mathcal{P}$ is related to the stereographic projection of the third Hopf fibration of the form
\begin{equation}
\mathcal{P}:  \mathbb{O}^{2} \xrightarrow{}   \mathbb{O}\mathbb{P}^{1},
\end{equation}
where $\mathbb{O}\mathbb{P}^{1}$ is the one dimensional octonionic projective space.
An important characteristic of the stereographic projection  (\ref{confo2}), (or third Hopf fibration in \cite{rrd, phdm}), is its sensitivity to entanglement.
To give it a more concrete meaning, we must express the entanglement of three-qubit state (\ref{threequbit}) in terms of concurrence measure introduced by Akhtarshenas \cite{Akhtarshenas}. To this end we assume general form of bipartite pure state
 \be\label{q}
\left| \psi  \right\rangle_{AB}  = \sum\limits_{i = 1}^{N_1 } {\sum\limits_{j = 1}^{N_2 } {a_{ij} \left| {i \rangle \otimes |j } \right\rangle }},
\ee
which  the norm of concurrence vector in terms of the $a_{ij}$ reads as
 \be\label{cc}
C^{AB} = 2  \sqrt {\sum\limits_{i < j}^{N_1 } {\sum\limits_{k < l}^{N_2 } {\left| {a_{ik} a_{jl} - a_{il} a_{jk} } \right|^2 } } },
\ee
If we  take the first qubit as partition $A$ and the last two qubits as partition $B$ in three-qubit state (\ref{threequbit}), then
its concurrence can be calculated in terms of coefficients $t_{i}$ as
\be\label{1(23)}
\begin{array}{l}
C^{1(23)}=2(|t_{0}t_{5}-t_{1}t_{4}|^{2}+|t_{0}t_{6}-t_{2}t_{4}|^{2}+|t_{0}t_{7}-t_{3}t_{4}|^{2}\\
\quad\quad \quad\quad+|t_{1}t_{6}-t_{2}t_{5}|^{2}+|t_{1}t_{7}-t_{3}t_{5}|^{2}+|t_{2}t_{7}-t_{3}t_{6}|^{2})^{\frac{1}{2}}.
 \end{array}
\ee
Now if $C^{1(23)}=0$, i.e. first qubit is factorized from the last two qubits, then the terms  $S_{1},S_{2}$ and $S_{3}$ in stereographic projection (\ref{confo2}) vanish meaning that the separable states is projected  to complex plane. Note that   the inverse of this  statement is not true, i.e. there are points  on  complex plane that  have no separable counterpart at all.
\par
Following the same logic as marshalled in section 2 for  two-qubit states, we consider the general form of local unitary transformation that act on three-qubit state (\ref{threequbit}) as
\begin{equation}
| \psi \rangle  \to | \psi \rangle ' = {A_1} \otimes {A_2} \otimes {A_3}| \psi  \rangle.
\end{equation}
where all ${A_i}$ belong to the group $ SU(2)$ and they can be parameterized as
\begin{equation}\label{lut}
{A_{i}} = \left( {\begin{array}{*{20}{c}}
  {{a_{i}}}&{{b_{i}}} \\
  { - {{\bar b}_{i}}}&{{{\bar a}_{i}}}
\end{array}} \right),  \quad\quad |a_{i}|^{2}+|b_{i}|^{2}=1\ .
\end{equation}
 Writing  three-qubit state (\ref{threequbit}) in the aforementioned  quaternion and octonion forms in Eqs. (\ref{quater1}) and (\ref{octabit}), we can  perform  local unitary transformation as follows
\begin{equation}\label{qv}
 | \psi \rangle '_{q} ={A_1} \otimes {A_2} [ | \psi  \rangle_{q} \mathcal{A}_{3}^{(q)}]= {A_1} \otimes {A_2} [ | \psi  \rangle_{q}(a_{3}-\bar{b}_{3}e_{2})],
\end{equation}
\begin{equation}\label{qqq}
| \psi \rangle '_{o} = {A_1}\{ [ | \psi  \rangle_{o}(\mathcal{A}_{3}^{(q)})](\mathcal{A}_{2}^{(o)})\},
\end{equation}
where the octonionic representation  $\mathcal{A}_{2}^{(o)}=a_{2}-\bar{b}_{2}e_{4}$ is defined analogous to Eq. (\ref{iso1}) except that  $e_{2}$ is replaced by $e_{4}$.
 We now  make the octonionic stereographic projection which includes the local unitary groups as
\begin{align}\label{cmo5}
 &\mathcal{P}((|{\psi}\rangle'_{o})=\frac{1}{|o'_{2}|^{2}}
\frac{(a_{1} \tilde o_{1} +b_{1}\tilde o_{2})( [\mathcal{A}_{3}^{(q)} \{\mathcal{A}_{2}^{(o)}\bar{\mathcal{A}}_{2}^{(o)}\} \bar{\mathcal{A}}_{3}^{(q)}])(-\bar {\tilde o}_{1} b+\bar {\tilde o}_{2}a)}
{|(-\bar{b_{1}}[\{ {\tilde o}_{1}\mathcal{A}_{3}^{(q)}\} \mathcal{A}_{2}^{(o)}])+(\bar{a}_{1}[\{\tilde o_{2}(\mathcal{A}_{3}^{(q)}\}\mathcal{A}_{2}^{(o)}])|^{2}},
\end{align}
where $o'_{2}=-\bar b_{1}[\{o_{1}\mathcal{A}_{3}^{(q)}\}\mathcal{A}_{2}^{(o)}]+\bar a_{1}[\{o_{2}\mathcal{A}_{3}^{(q)}\}\mathcal{A}_{2}^{(o)}]$. The definition (\ref{cmo5})  reduces to Eq. (\ref{confo2}), whenever we take $A_{1},A_{2}$ and $A_{3}$ as identity operator $I$.
\par
Finally, we define the octonionic M\"{o}bius transformations as
 \begin{align}\label{OMT}
&\mathcal{F}_{_{B}}(o)
 :=\frac{(a_{1} o +b_{1})( [\mathcal{A}_{3}^{(q)} \{\mathcal{A}_{2}^{(o)}\bar{\mathcal{A}}_{2}^{(o)}\} \bar{\mathcal{A}}_{3}^{(q)}])(-\bar o b+a)}{|-\bar b_{1}[\{o\mathcal{A}_{3}^{(q)}\} \mathcal{A}_{2}^{(o)}] +\bar a_{1} \{\mathcal{A}_{3}^{(q)} \mathcal{A}_{2}^{(o)}\}|^{2}},
\end{align}
where $o=\mathcal{P}(|{\psi}\rangle_{o})$ is stereographic projection of the initial state (\ref{threequbit}).
This unique choice of  octonionic M\"{o}bius transformation is based on the implicit fact that we treat the space of octonionic spinors as a right module (multiplication by quaternions and octonions from the right).
Regarding above considerations, we finally state our main result  in the commutative diagram  as
\[\begin{CD}
{\mathcal{H}}_{8}^{\mathbb{C}}  @>{\mathcal{Q}}>>  {\mathcal{H}}_{4}^{\mathbb{Q}}   @>{\mathcal{O}}>>  {\mathcal{H}}_{2}^{\mathbb{O}}    @>{\mathcal{P}}>> \mathbb{\widetilde{O}} \\
@V{B}VV              @VV{\mathbb{Q}B}V                    @VV{\mathbb{O}B}V       @VV{\mathcal{F}_{_{B}}}V\\
{\mathcal{H}}_{8}^{\mathbb{C}}  @>{\mathcal{Q}}>>  {\mathcal{H}}_{4}^{\mathbb{Q}}   @>{\mathcal{O}}>>  {\mathcal{H}}_{2}^{\mathbb{O}}    @>{\mathcal{P}}>> \mathbb{\widetilde{O}}
\end{CD}\]
In the present case commutativity of the diagram is equivalent to the commutativity relation
 (for more detail see Appendix B)
\begin{equation}\label{eq1}
\mathcal{P}\mathcal{O}\mathcal{Q}(B)|\psi\rangle
=\mathcal{P}\mathcal{O}(\mathbb{Q}B)\mathcal{Q}|\psi\rangle
=\mathcal{P}(\mathbb{O}B)\mathcal{O}\mathcal{Q}|\psi\rangle
=\mathcal{F}_{_{B}}\mathcal{P}\mathcal{O}\mathcal{Q}|\psi\rangle,
\end{equation}
Before concluding this section, let
us make some preliminary remarks on the general properties of this diagram. As it is evident from the last equality, we have
$\mathcal{P}(\mathbb{O}B)|\psi\rangle_{o}
=\mathcal{F}_{_{B}}\mathcal{P}|\psi\rangle_{o},$ i.e. the octonion stereographic projection $\mathcal{P}$ intertwines between $(\mathbb{O}B)$ and octonionic M\"{o}bius transformation $\mathcal{F}_{_{B}}$.
As mentioned earlier in this section, the terms $S_{1},S_{2}$ and $S_{3}$ are entanglement sensitive, therefore we expect that they become invariant  under local unitary transformations, i.e. $S'_{i}=S_{i}$ hold for $i=1,2,3$ (for aprecise proof, see Appendix B).

\section{Conclusion}
In summary, we have considered some  geometrical aspects  of one, two and three-qubit pure states under local unitary transformation.
In this investigation we saw that the stereographic projection $\mathcal{P}$ plays a crucial role. For one-qubit states, the mapping is one to one correspondence between Bloch sphere and complex projective space $\mathbb{\widetilde{C}}$. In addition, $\mathcal{P}$ intertwines between unitary group  $SU(2)$  and its corresponding complex M\"{o}bius transformation. As an example we have discussed  the time evolution of one-qubit  state under  Hadamard like Hamiltonian in more detail. Considering  two-qubit states we observe the following features: A rather generalized  intertwining   stereographic projection   $\mathcal{P}$, between  global group  $Sp(2)$  and its corresponding quaternionic M\"{o}bius transformation, is easy to construct, but in general
 it is not completely equivalent to work with
the two-qubit states due to the non-commutativity of quaternion numbers in its definition. However, for local unitary transformation $SU(2)\otimes SU(2)$ there is unique quaternionic stereographic projection $\mathcal{P}$ and M\"{o}bius transformation, both leave the concurrence term invariant and fulfill the Eq.  (\ref{ww}), i.e. we have a commuting diagram.
The geometry of the three-qubit  is generally richer than that of two-qubit states.
More generally, the octonionic representation of three-qubit states
cannot be transformed under a  global group  which places it outside the line of  treatments of one and two-qubit cases. This is due to the fact that, octonion numbers are not only noncommutative but also they are non-associative.
For a different task, however, the construction  of octonionic stereographic projection $\mathcal{P}$ and M\"{o}bius transformation under local unitary group turns out to be possible. As in the case of two-qubits, both  octonionic stereographic projection and M\"{o}bius transformation are entanglement sensitive, i.e. there are terms that is invariant under local unitary transformation $SU(2)\otimes SU(2)\otimes SU(2)$  and have something to do with concurrence measure. In all of these construction, we have insisted  on commutativity of diagrams.

\vspace{1cm} \setcounter{section}{0}
 \setcounter{equation}{0}
 \renewcommand{\theequation}{A-\arabic{equation}}
\newpage
\textbf{Appendix A: Quaternion and Octonion}\\
The quaternion skew-field $\mathbb{Q}$ is an associative algebra of
rank 4 on real number $\mathbb{R}$ whose every element can be written as
\begin{equation}
q=\sum\limits_{i = 0}^3 {{x_i}{e_i}} ,\quad \quad {x_i} \in \mathbb{R},
\end{equation}
where $e_{0}=1$ and  $e_{1},e_{2},e_{3}$ satisfy
\begin{equation}
e_{1}^{2}=e_{2}^{2}=e_{3}^{2}=-1, \quad e_{1}e_{2}=-e_{2}e_{1}=e_{3}, \quad e_{2}e_{3}=-e_{3}e_{2}=e_{1}, \quad e_{3}e_{1}=-e_{1}e_{3}=e_{2}.
\end{equation}
It can also be defined equivalently, using the complex numbers
$z_{1}=x_{0}+x_{1}e_{1}$  and
 $z_{2}=x_{2}+x_{3}e_{1} $ in the form  $q=z_{1}+z_{2}e_{2} $
 endowed with an involutory anti-automorphism (conjugation) such as
\begin{equation}
q=z_{1}+z_{2}e_{2} \in \mathbb{C}\oplus\mathbb{C}e_{2}\ \longrightarrow \
\bar{q}=x_{0}-\sum\limits_{i = 1}^3 {{x_i}{e_i}}=\bar{z}_{1}-z_{2}e_{2}.
\end{equation}
 Note that quaternion multiplication is non-commutative so that
 $\overline{q_{1}q_{2}}={\overline{q_{2}}}\ {\overline{q_{1}}}$ and $
e_{2}z=\bar{z}e_{2}$. On the other hand a two dimensional quaternionic vector space $V$
defines a four dimensional complex vector space $\mathbb{C}V$ by
forgetting  scalar multiplication by non-complex quaternions
(i.e., those involving $e_{2}$). Roughly
speaking if $V$ has quaternionic dimension $2$, with basis
$|0\rangle_{q},|1\rangle_{q}$, then
$\mathbb{C}V$  has complex dimension 4, with basis
$\{|00\rangle,|01\rangle,|10\rangle,|11\rangle\}$.\\
\begin{figure}
\centerline{\includegraphics[width=11cm]{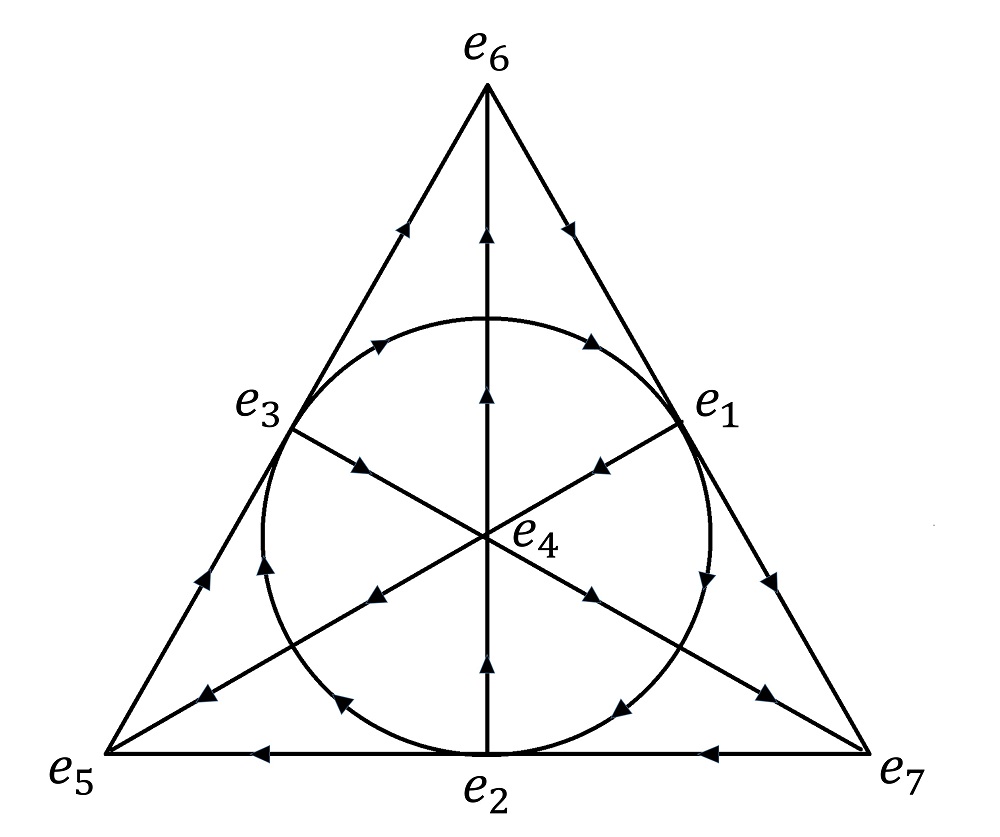}}
\caption{\label{appendix Fig.3}\small { Multiplication rule for octonion.} \label{appendix Fig.3}}
\end{figure}
The octonionic skew-field $\mathbb{O}$ is neither commutative nor  associative algebra of
rank 8 over $\mathbb{R}$ whose every element can be written as
\begin{equation}
o=\sum\limits_{i = 0}^7 {{x_i}{e_i}} \quad \quad ,{x_i} \in \mathbb{R}\quad \quad ,{e_0} = 1\quad \mathrm{and}  \quad e_i^2 =  - 1{\kern 1pt} {\kern 1pt} (i = 1,...,7).
\end{equation}
The multiplication of octonion is given by
\begin{equation}
{e_i}{e_j} =  - {\delta _{ij}}{e_0} + \sum\limits_{k = 1}^7 {{f_{ijk}}{e_k}} \quad \quad \mathrm{for} \quad \quad i,j = 1,2,...,7\ \ ,
\end{equation}
where the ${f_{ijk}}$ are totally anti-symmetric in $i, j$ and $k$
with values $1, 0,-1$ just as  Levi-Civita symbol. Moreover, $f_{ijk} = +1$ for $ijk= 123, 145, 246, 347,617, 725,536$. We can depict this fact graphically as in  (Fig. \ref{appendix Fig.3}) where the multiplication role can be read from orientation of arrows, for example:
\begin{equation}
e_{1}e_{2}=e_{3} , \quad \quad e_{2}e_{4}=e_{6}  , \quad \quad e_{7}e_{2}=e_{5}.
\end{equation}
The complex conjugate of a octonion is given by
\begin{equation}
\bar o=x_{0}-\sum\limits_{i = 1}^7 {{x_i}{e_i}}.
\end{equation}
 Any octonion number can be represented  in terms of four complex number $ z_{0}=x_{0}+x_{1}e_{1}$, $ z_{1}=x_{2}+x_{3}e_{1}$, $z_{2}=x_{4}+x_{5}e_{1}$ and $z_{3}=x_{6}+x_{7}e_{1}$ as
\begin{equation}
o=z_{0}+z_{1}e_{2}+(z_{3}+z_{3}e_{2})e_{4},
\end{equation}
and it's complex conjugate is given by
\begin{equation}
\bar o=\bar z_{0}-z_{1}e_{2}-(z_{3}+z_{3}e_{2})e_{4},
\end{equation}
or equivalently, in terms  of quaternion numbers $q_{1}=z_{0}+z_{1}e_{2}$  and  $q_{2}=z_{2}+x_{3}e_{2} $, we can write
\begin{align}
\begin{array}{c}
o=q_{1}+q_{2}e_{4}\ , \\
\bar o=\bar q_{1}-q_{2}e_{4}\ \\.
\end{array}
\end{align}
The multiplication of two octonions
\begin{align}
\begin{array}{c}
 o_{1}=z_{0}+z_{1}e_{2}+(z_{3}+z_{3}e_{2})e_{4}\ ,\\
   o_{2}=p_{0}+p_{1}e_{2}+(p_{3}+p_{3}e_{2})e_{4}\ , \\
\end{array}
\end{align}
using the multiplication rule of $e_{i}$, is an octonion
\begin{equation}
o_{3}=o_{1}o_{2}=s_{0}+s_{1}e_{2}+(s_{3}+s_{3}e_{2})e_{4} ,
\end{equation}
where $s_{i}, i=0,1,2,3$ are complex numbers and are given by
\begin{align}
\begin{array}{c}
  s_{0}=z_{0}p_{0}-z_{1}\bar p_{1}-z_{2}\bar p_{2}-\bar z_{3}p_{3}, \\
  s_{1}=z_{0}p_{1}+z_{1}\bar p_{0}+\bar z_{2} p_{3}- z_{3}\bar p_{2}, \\
  s_{2}=z_{0}p_{2}-\bar z_{1} p_{3}+z_{2} \bar p_{0}+ z_{3}\bar p_{1}, \\
  s_{3}=\bar z_{0} p_{3}+ z_{1} p_{2}-z_{2} \bar p_{1}+ z_{3}\bar p_{0}.
\end{array}
\end{align}
Every non-zero ($p\in \mathbb{Q}$ or $\mathbb{O}$)  is invertible, and the unique inverse is
given by $p^{-1}=\frac{1}{|p|^2}\bar{p}$
where the quaternion or octonionic  norm $|p|$ is defined by $
|p|^2=p\bar{p}.$
The norm of two  quaternions or octonions  $p_{1}$ and $p_{2}$ satisfies $
|p_{1}p_{2}| = |p_{2}p_{1}| = |p_{1}||p_{2}|.$
 Note that octonion multiplication is non-commutative and non-associative that  is
 $(o_{1}o_{2})o_{3}\neq o_{1}(o_{2}o_{3})$.
 On the other hand a two dimensional octonionic vector space $V$
defines a four dimensional quaternionic vector space $\mathbb{Q}V$ by
forgetting scalar multiplication by octonion
(i.e., those involving $e_{4}$). Roughly
speaking if $V$ has octonionic dimension $2$, with basis
$\{|0\rangle_{o},|1\rangle_{0}\}$, then
$\mathbb{Q}V$  has quaternion dimension 4, with basis
$\{|00\rangle_{q},|01\rangle_{q},|10\rangle_{q},|11\rangle_{q}\}$ and $\mathbb{C}V$  has complex dimension 8, with basis
$\{|000\rangle,|001\rangle,|010\rangle,|011\rangle,|100\rangle,|101\rangle,|110\rangle,|111\rangle\}$.\\
\par
\setcounter{equation}{0}
 \renewcommand{\theequation}{B-\arabic{equation}}
\par
\textbf{Appendix B: Calculating }\\
In this appendix we calculate the terms appearing in the main
commutative diagram related to three-qubit pure state.
\par
\textbf{Calculating $\mathcal{P}\mathcal{O}\mathcal{Q}B|\psi\rangle$}\\
It is convenient to start with the first statement
in equation(\ref{eq1}),
\begin{equation}\label{path1}
\mathcal{P}\mathcal{O}\mathcal{Q}B|\psi\rangle=\mathcal{P}\mathcal{O}\mathcal{Q}|\psi\rangle'=\mathcal{P}\mathcal{O}|\psi\rangle'_{q}=\mathcal{P}|\psi\rangle'_{o}  ,
\end{equation}
where $|\psi\rangle'_{o}=o'_{1}|0\rangle_{o}+o'_{2}|1\rangle_{o}$ and $o'_{1}$ and $o'_{2}$ are results of the action of local unitary transformation $B=A_{1}\otimes A_{2}\otimes A_{3}$ on the general three-qubit pure state (\ref{threequbit}) followed by the map $\mathcal{O}\mathcal{Q}$ as
\begin{align}\label{1o'}
&{ o}'_{1}=t_{0}'+t_{1}'e_{2}+(t_{2}'+t_{3}'e_{2})e_{4}= \quad \quad \quad \quad \quad \quad \quad \quad \quad \quad \quad \quad \quad \quad \quad \quad \quad \quad \quad \quad  \quad \quad \quad \quad \quad \quad \quad \quad \quad \quad \quad \quad \quad \quad \quad \quad \quad\quad \quad \quad \quad\quad \quad \quad \quad \nonumber\\
&\quad (t_{0}{a_{1}}{a_{2}}{a_{3}}+t_{1}{a_{1}}{a_{2}}{b_{3}}+t_{2}{a_{1}}{b_{2}}\bar{a_{3}}+t_{3}{a_{1}}{b_{2}}{b_{3}}+t_{4}{b_{1}}{a_{2}}{a_{3}}+t_{5}{b_{1}}{a_{2}}{b_{3}}+t_{6}{b_{1}}{b_{2}}{a_{3}}+t_{7}{b_{1}}{b_{2}}{b_{3}})\nonumber\\
&+(-t_{0}{a_{1}}\bar{a_{2}}\bar{b_{3}}+t_{1}{a_{1}}\bar{a_{2}}\bar{a_{3}}-t_{2}{a_{1}}{b_{2}}\bar{b_{3}}+t_{3}{a_{1}}{b_{2}}\bar{a_{3}}-t_{4}{b_{1}}\bar{a_{2}}\bar{b_{3}}+t_{5}{b_{1}}{a_{2}}{a_{3}}-t_{6}{b_{1}}{b_{2}}\bar{b_{3}}+t_{7}{b_{1}}{b_{2}}\bar{a_{3}})e_{2}\nonumber\\
&+(-t_{0}{a_{1}}\bar{b_{2}}{a_{3}}-t_{1}{a_{1}}\bar{b_{2}}{b_{3}}+t_{2}{a_{1}}\bar{a_{2}}{a_{3}}+t_{3}{a_{1}}\bar{a_{2}}{b_{3}}-t_{4}{b_{1}}\bar{b_{2}}{a_{3}}-t_{5}\bar{b_{1}}\bar{b_{2}}\bar{b_{3}}+t_{6}{b_{1}}\bar{a_{2}}{a_{3}}+t_{7}{b_{1}}\bar{a_{2}}{b_{3}})e_{4}\nonumber\\
&+([t_{0}{a_{1}}{b_{2}}\bar{b_{3}}-t_{1}{a_{1}}\bar{b_{2}}\bar{a_{3}}-t_{2}{a_{1}}\bar{a_{2}}\bar{b_{3}}+t_{3}{a_{1}}\bar{a_{2}}\bar{a_{3}}+t_{4}{b_{1}}{b_{2}}\bar{b_{3}}-t_{5}{b_{1}}\bar{b_{2}}\bar{a_{3}}-t_{6}{b_{1}}\bar{a_{2}}\bar{b_{3}}+t_{7}{b_{1}}\bar{a_{2}}\bar{a_{3}}]e_{2})e_{4},\nonumber\\
\end{align}
\begin{align}\label{2o'}
&{ o}'_{2}=t_{4}'+t_{5}'e_{2}+(t_{6}'+t_{7}'e_{2})e_{4}= \quad \quad \quad \quad \quad \quad \quad \quad \quad \quad \quad \quad \quad \quad \quad \quad \quad \quad \quad \quad \quad \quad \quad \quad  \quad \quad \quad\nonumber\\
&\quad(-t_{0}\bar{b_{1}}{a_{2}}{a_{3}}-t_{1}\bar{b_{1}}{a_{2}}{b_{3}}-t_{2}\bar{b_{1}}{b_{2}}{a_{3}}-t_{3}\bar{b_{1}}{b_{2}}{b_{3}}+t_{4}\bar{a_{1}}{a_{2}}{a_{3}}+t_{5}\bar{a_{1}}{a_{2}}{b_{3}}+t_{6}\bar{a_{1}}{b_{2}}{a_{3}}+t_{7}\bar{a_{1}}{b_{2}}{b_{3}})\nonumber\\
&+(t_{0}\bar{b_{1}}\bar{a_{2}}\bar{b_{3}}-t_{1}\bar{b_{1}}\bar{a_{2}}\bar{a_{3}}+t_{2}\bar{b_{1}}{b_{2}}\bar{b_{3}}-t_{3}\bar{b_{1}}r{b_{2}}\bar{a_{3}}-t_{4}\bar{a_{1}}\bar{a_{2}}\bar{b_{3}}+t_{5}\bar{a_{1}}\bar{a_{2}}\bar{a_{3}}-t_{6}\bar{a_{1}}{b_{2}}\bar{b_{3}}+t_{7}\bar{a_{1}}{b_{2}}\bar{a_{3}})e_{2}\nonumber\\
&+(t_{0}\bar{b_{1}}\bar{b_{2}}{a_{3}}+t_{1}\bar{b_{1}}\bar{b_{2}}{b_{3}}-t_{2}\bar{b_{1}}\bar{a_{2}}{a_{3}}-t_{3}\bar{b_{1}}\bar{a_{2}}{b_{3}}-t_{4}\bar{a_{1}}\bar{b_{2}}{a_{3}}-t_{5}\bar{a_{1}}\bar{b_{2}}{b_{3}}+t_{6}\bar{a_{1}}\bar{a_{2}}{a_{3}}+t_{7}\bar{a_{1}}\bar{a_{2}}{b_{3}})e_{4}\nonumber\\
&+([t_{0}\bar{b_{1}}\bar{b_{2}}\bar{b_{3}}+t_{1}\bar{b_{1}}\bar{b_{2}}\bar{a_{3}}+t_{2}\bar{b_{1}}\bar{a_{2}}\bar{b_{3}}-t_{3}\bar{b_{1}}\bar{a_{2}}\bar{a_{3}}+t_{4}\bar{a_{1}}\bar{b_{2}}\bar{b_{3}}-t_{5}\bar{a_{1}}\bar{b_{2}}\bar{a_{3}}-t_{6}\bar{a_{1}}\bar{a_{2}}\bar{b_{3}}+t_{7}\bar{a_{1}}\bar{a_{2}}\bar{a_{3}}]e_{2})e_{4}, \quad \quad
\end{align}
According to the relation $\mathcal{O}|\psi\rangle'_{o}=(\mathcal{O}B)\mathcal{O}|\psi\rangle_{o}$, the $o'_{1}$ and $o'_{2}$ can be factorized   as
\begin{align}\label{o'}
\begin{array}{c}
o'_{1}=a_{1}[\{o_{1}\mathcal{A}_{3}^{(q)}\}\mathcal{A}_{2}^{(o)}]+b_{1}[\{o_{2}\mathcal{A}_{3}^{(q)}\}\mathcal{A}_{2}^{(o)}]\ ,\\
  o'_{2}=-\bar b_{1}[\{o_{1}\mathcal{A}_{3}^{(q)}\}\mathcal{A}_{2}^{(o)}]+\bar a_{1}[\{o_{2}\mathcal{A}_{3}^{(q)}\}\mathcal{A}_{2}^{(o)}]\ , \\
\end{array}
\end{align}
where $\mathcal{A}_{3}^{(q)}$ and $\mathcal{A}_{2}^{(o)}$ are quaternion and octonionic form of unitary transformation $A_{3}$ and $A_{2}$, introduced in (\ref{iso1}) and (\ref{qqq}), respectively.
\par
\textbf{ Calculating $\mathcal{P}\mathcal{O}(\mathbb{Q}B)\mathcal{Q}|\psi\rangle$}\\
We can proceed another approach to understand more about three-qubit entangled pure state. Unlike in definition of $\mathbb{Q}B$, in order to correctly represent the complex transformation on the three-qubit pure state, the   transformation $B$  on the two-quaterbit should be represented by left action of the $4 \times 4$ matrix $A_{1} \otimes  A_{2}$ and right multiplication with the quaternion $\mathcal{A}_{3}^{(q)}$ as in equation (\ref{qv}), hence  applying the $\mathbb{Q}B$ on a two-quaterbit $|\psi \rangle_{q}$ one can get
\begin{equation}\label{p2}
A_{1} \otimes  A_{2}[|\psi\rangle \mathcal{A}_{3}^{(q)}]=\left( {\begin{array}{*{20}{c}}
  {t{'_0} + t{'_1}{e_2}} \\
  {t{'_2} + t{'_3}{e_2}} \\
  {t{'_4} + t{'_5}{e_2}} \\
  {t{'_6} + t{'_7}{e_2}}
\end{array}} \right),
\end{equation}
where $t'_{0},...,t'_{7}$ are the same forms as in equation (\ref{1o'}). It is clear that the operation  $\mathcal{O}$ leads to equation (\ref{o'}), hence the relevant part of the crucial diagram (the first quadrangle) is commutative, i.e.,
\begin{equation}\label{p2}
\mathcal{O}\mathcal{Q}B|\psi\rangle
=\mathcal{O}(\mathbb{Q}B)\mathcal{Q}|\psi\rangle,
\end{equation}
and it is clear that applying the octonionic stereographic projection on the both side of the above equation lead to the first equality in equation  (\ref{eq1}).
\par
\textbf{ Calculating $\mathcal{P}(\mathbb{O}B)\mathcal{O}\mathcal{Q}|\psi\rangle$}\\
We mentioned that the octonionic form of local unitary transformation  should be represented by left action of the $2 \times 2$ matrix $(A_{1}\in SU(2))$ and right multiplication with the octonion $\mathcal{A}_{3}^{(q)}, \mathcal{A}_{2}^{(o)}$  keeping in mind that the ordering is as Eq. (\ref{o'}). Now, applying the $(\mathbb{O}B)$ on a octabit one can get
\begin{equation}\label{p2}
{A_1}\{ [ | \psi  \rangle_{o}\mathcal{A}_{3}^{(q)}]\mathcal{A}_{2}^{(o)}\}=
\left( {\begin{array}{*{20}{c}}
  {{a_1}}&{b_{1}} \\
 {-\bar{b_{1}}}&{{a_1}}
\end{array}} \right)\left( {\begin{array}{*{20}{c}}
  {\{ {o_1}\mathcal{A}_{3}^{(q)}\}\mathcal{A}_{2}^{(o)}} \\
  {\{ {o_2}\mathcal{A}_{3}^{(q)}\} \mathcal{A}_{2}^{(o)}}
\end{array}} \right)=\left( {\begin{array}{*{20}{c}}
  {o{'_1}} \\
  {o{'_2}}
\end{array}} \right),
\end{equation}
where $o{'_1}$ and $o{'_2}$ have the same forms as in equation  (\ref{o'}). Altogether, we obtain
\begin{equation}\label{eq2}
\mathcal{O}\mathcal{Q}(\mathbb{C}B)|\psi\rangle
=\mathcal{O}(\mathbb{Q}B)\mathcal{Q}|\psi\rangle
=(\mathbb{O}B)\mathcal{O}\mathcal{Q}|\psi\rangle,
\end{equation}
Now, applying the octonionic stereographic projection $\mathcal{P}$ on above equation lead to the second equality in (\ref{eq1}), or more precisely we have
\begin{equation}\label{c1}
\mathcal{P}(|\psi\rangle'_{o})=\frac{1}{|o'_{2}|^{2}}[a_{1}b_{1}(\tilde o_{2}\bar{\tilde o}_{2}-o_{1}\bar o_{1})+(a_{1}\frac{o_{1}\bar{\tilde o}_{2}}{|o_{2}|^{2}}a_{1}-b_{1}\frac{{\tilde o_{2}}\bar o_{1}}{|o_{2}|^{2}}b_{1})],
\end{equation}
where we used  the relations
$\tilde o_{2}\bar{\tilde o}_{2}-o_{1}\bar o_{1}=|o_{2}|^{2}(1-o\bar o)$)  and $o=\mathcal{P}(|{\psi}\rangle_{o})=\frac{\tilde o_{1}\bar{\tilde o}_{2}}{|o_{2}|^{2}}$ to get
\begin{align}\label{c2}
 &\mathcal{P}(o'_{1}|0\rangle_{o}+o'_{2}|1\rangle_{o})=\frac{|o_{2}|^{2}}{|o'_{2}|^{2}}(a_{1}b_{1}(1-o\bar o)+a_{1}oa_{1}-b_{1}\bar o b_{1}) \nonumber\\
 & =\frac{|o_{2}|^{2}}{|o'_{2}|^{2}}[S'_{0}+S'_{1}e_{2}+(S'_{2}+S'_{3}e_{2})e_{4}],
\end{align}
where
\begin{align}\label{c3}
\begin{array}{c}
S'_{0}=[a_{1}b_{1}(1-|S_{0}|^2-|S_{1}|^2-|S_{2}|^2-|S_{3}|^2)+(a_{1}^{2}S_{0}-b_{1}^{2}\bar{S_{0}})+S_{1}e_{2}\ ,\\
S'_{1}=S_{1}\ ,\quad \quad \quad \quad \quad \quad \quad \quad \quad \quad \quad \quad \quad \quad \quad \quad\quad \quad \quad  \quad \quad \quad \quad\quad \quad  \\
S'_{2}=S_{2}\ ,\quad \quad \quad \quad \quad \quad \quad \quad \quad \quad \quad \quad \quad \quad \quad \quad \quad \quad \quad \quad \quad \quad \quad \quad\quad  \\
 S'_{3}=S_{3}\ ,\quad \quad \quad \quad \quad \quad \quad \quad \quad \quad \quad \quad \quad \quad \quad \quad \quad \quad \quad \quad \quad \quad \quad \quad\quad   \\
\end{array}
\end{align}
and $S_{0}, S_{1}, S_{2}$ and $S_{3}$ are complex number that we introduce in (\ref{cmot})
\par
\textbf{ Calculating $\mathcal{F}_{_{B}}\mathcal{P}\mathcal{O}\mathcal{Q}|\psi\rangle$}\\
So far we have  shown that the first two equalities in equation (\ref{eq1})
holds. We will henceforth focus on the role of octonionic M\"{o}bius transformation.
 Using the linear map $\mathcal{Q}$ and $\mathcal{O}$ together with octonionic stereographic projection $\mathcal{P}$ on a  three-qubit pure state in equation (\ref{threequbit}) yields
\begin{equation}
\mathcal{P}\mathcal{O}\mathcal{Q}|\psi\rangle=\mathcal{P}\mathcal{O}|\psi\rangle_{q}=\mathcal{P}|\psi\rangle_{o}=\tilde o_{1}\tilde o^{-1}_{2}=\frac{1}{|o_{2}|^2 }(S_{0}+S_{1}e_{2}+(S_{2}+S_{3}e_{2})e_{4}).
\end{equation}
Furthermore, this point is mapped under the action of the octonionic M\"{o}bius transformation in equation(\ref{OMT}) as follows
\begin{align}\label{cmot2}
 &\mathcal{F}_{B}(\frac{1}{|o_{2}|^2}(S_{0}+S_{1}e_{2}+(S_{2}+S_{3}e_{2})e_{4})\nonumber\\
 &= \frac{1}{|o_{2}|^2}[a_{1}b_{1}(1-|S_{0}|^2-|S_{1}|^2-|S_{2}|^2-|S_{3}|^2)+(a_{1}^{2}S_{0}-b_{1}^{2}\bar{S_{0}})+S_{1}e_{2}+(S_{2}e_{4}+S_{3}e_{2})e_{4})]\nonumber\\
 & =\frac{1}{|o_{2}|^2}[S'_{0}+S'_{1}e_{2}+(S'_{2}e_{4}+S'_{3}e_{2})e_{4}],
\end{align}
in which  the term $S_{1}e_{2}+S_{2}e_{4}+(S_{3}e_{2})e_{4}$ is invariant.
 Remembering that the entanglement between the first qubit and the last two qubits of the state (\ref{threequbit}) is given by concurrence measure  (\ref{1(23)}), we have  $S_{1}=S_{2}=S_{3}=0$ for separable three-qubit states. For completeness we mention that the first qubit   lives on the base space of the Hopf fibration while the other two-qubit    lives on the fiber space of the Hopf map.
 On the other hand, since there is, in general, a symmetry in choosing the partitions of a three-qubit  state,
 we could have started from
  $|{\psi}\rangle_{2,(13)}$ or $|{\psi}\rangle_{3,(12)}$ (instead of $|{\psi}\rangle_{1,(23)}$), hence there are  three kind of  octonionification and subsequently   three stereographic projections. Two others are
\begin{align}\label{octabit2}
 &|{\psi}\rangle_{o'}=m_{1}|{ 0 }\rangle_{o'}+m_{2}|{ 1 }\rangle_{o'}\nonumber\\
 &|{\psi}\rangle_{o''}=n_{1}|{ 0 }\rangle_{o''}+n_{2}|{ 1 }\rangle_{o''},
\end{align}
where $m_{1}, m_{2}, n_{1}, n_{2}$ are octonion numbers
\begin{align}\label{o2}
 &m_{1}=t_{0}+t_{1}e_{2}+(t_{4}+ t_{5}e_{2})e_{4},  \nonumber\\
 &m_{2}=t_{2}+t_{3}e_{2}+(t_{6}+ t_{7}e_{2})e_{4},  \nonumber\\
 &n_{1}=t_{0}+t_{2}e_{2}+(t_{4}+ t_{6}e_{2})e_{4}, \nonumber\\
 & n_{2}=t_{1}+t_{3}e_{2}+(t_{5}+ t_{7}e_{2})e_{4}.
\end{align}
Acting the stereographic projection $\mathcal{P}$ on two states $|{\psi}\rangle_{o'}$ and $|{\psi}\rangle_{o''}$ yields
\begin{align}\label{canformal23}
 &M=( t_{0}\bar{t_{2}}+ t_{1}\bar{t_{3}}+ t_{4}\bar{t_{6}}+ t_{3}\bar{t_{7}})+( t_{1}t_{2} - t_{0}t_{3}+\bar{t_{5}}\bar{t_{6}}-\bar{t_{4}}\bar{t_{7}})e_{2}  \nonumber\\
 &+[(  t_{4}t_{2} - t_{0}t_{6}+\bar{t_{1}}\bar{t_{7}}-\bar{t_{5}}\bar{t_{3}})+( t_{4}t_{3} - t_{1}t_{6}+\bar{t_{5}}\bar{t_{2}}-\bar{t_{0}}\bar{t_{7}})e_{2}]e_{4}  \nonumber\\
 &N=( t_{0}\bar{t_{1  }}+ t_{2  }\bar{t_{3  }}+ t_{4  }\bar{t_{5  }}+ t_{6  }\bar{t_{7}})+(  t_{2  }t_{1  } - t_{0}t_{3  }+\bar{t_{6  }}\bar{t_{5  }}-\bar{t_{4  }}\bar{t_{7}})e_{2} \nonumber\\
 &+[( t_{4  }t_{1  } - t_{0}t_{5  }+\bar{t_{2  }}\bar{t_{7}}-\bar{t_{6  }}\bar{t_{3  }})+( t_{4  }t_{3  } - t_{2  }t_{5  }+\bar{t_{6  }}\bar{t_{1  }}-\bar{t_{0}}\bar{t_{7}})e_{2}]e_{4}
\end{align}
respectively. Corresponding to each octonification there are concurrence measures  which relate the entanglement between one qubit with  other two-qubit i.e.
\begin{equation}\label{2}
\begin{array}{l}
C^{2(13)}=2(|t_{0}t_{3}-t_{1}t_{2}|^{2}+|t_{0}t_{6}-t_{2}t_{4}|^{2}+|t_{0}t_{7}-t_{2}t_{5}|^{2}\\
\quad\quad \quad\quad  +|t_{1}t_{6}-t_{3}t_{4}|^{2}+|t_{1}t_{7}-t_{3}t_{5}|^{2}+|t_{4}t_{7}-t_{5}t_{6}|^{2})^{\frac{1}{2}},\\
C^{3(12)}=2(|t_{0}t_{3}-t_{1}t_{2}|^{2}+|t_{0}t_{5}-t_{1}t_{4}|^{2}+|t_{0}t_{7}-t_{1}t_{6}|^{2}\\
\quad\quad \quad\quad +|t_{2}t_{5}-t_{3}t_{4}|^{2}+|t_{2}t_{7}-t_{3}t_{6}|^{2}+|t_{4}t_{7}-t_{5}t_{6}|^{2})^{\frac{1}{2}}.\\
 \end{array}
\end{equation}
According the above concurrence,  the results $M$ and $N$, which arise from  projecting of the states $|{\psi}\rangle_{o'}$ and $|{\psi}\rangle_{o''}$, are also entanglement sensitive.
\par
\textbf{Acknowledgments}\\
The authors also acknowledge the support from the Mohaghegh Ardabili University.


\end{document}